\documentclass[a4paper,fleqn]{cas-sc}
\usepackage[numbers,sort&compress]{natbib} 
\pdfoutput=1 
 
\usepackage{overpic} 

\usepackage{threeparttable}

\def\tsc#1{\csdef{#1}{\textsc{\lowercase{#1}}\xspace}}
\tsc{WGM}
\tsc{QE}
\tsc{EP}
\tsc{PMS}
\tsc{BEC}
\tsc{DE}

\begin{document}
\let\WriteBookmarks\relax
\def\floatpagepagefraction{1}
\def\textpagefraction{.001}

\shorttitle{Kriging and Neural Network models for pressure losses}

\shortauthors{S Li et~al.}

\title [mode = title]{Kriging and Neural Network models for pressure losses across perforated plates}

\author[1]{Shuai Li}

\cormark[1]

\ead{shuai.li@chalmers.se.}

\credit{Conceptualization, Methodology, Software, Validation, Formal analysis, Investigation, Data curation, Writing - original draft, Writing - review \& editing, Visualization}

\address[1]{Division of Fluid Dynamics, Department of Mechanical Engineering, Chalmers University of Technology, SE-412 96 Gothenburg, Sweden}


\cortext[cor1]{Corresponding author.}

\begin{abstract}
In this paper, two novel data-driven models based on Kriging and neural networks (NN) are proposed to predict pressure losses across perforated plates with circular perforations in turbulent flows. The models are developed using two sets of experimental data available in the literature. The predictive performance of the proposed models is assessed and compared against widely used empirical formulae. It is found that the proposed models consistently outperform existing empirical models for most perforated plate configurations contained in the experimental datasets. Besides, the predicted pressure losses generally show good agreement with experimental measurements, demonstrating that data-driven approaches based on Kriging and NN provide a feasible framework for modelling pressure losses across perforated plates. Overall, both approaches are promising, despite being trained on a relatively limited amount of experimental data, owing to the scarcity of measurements reported in the literature. To demonstrate the applicability of the proposed models in numerical simulations, two-dimensional channel flows are simulated using the Reynolds-averaged Navier–Stokes (RANS) equations, in which the new pressure-loss models are implemented as a source term in the momentum equations. The RANS predictions are insensitive to mesh resolution and the number of cells used to represent the perforated plate, and they agree well with the model predictions, confirming the applicability of the proposed approaches to practical computational fluid dynamics (CFD) simulations.
\end{abstract}


\ExplSyntaxOn
\keys_set:nn { stm / mktitle } { nologo }
\ExplSyntaxOff

\maketitle

\section{Introduction}

Perforated plates are encountered in a wide range of engineering applications, including thermal, mechanical, chemical, civil, nuclear, ocean, and aerospace engineering, due to their ability to control flow distribution, enhance mixing, and regulate pressure losses. To name a few examples, perforated plates are widely used in the design of heat transfer devices \cite{ahmimache2022heat, lee2002heat, mcmahon1950perforated, shevyakova1983study, kutscher1994heat, white2010experimentally, tomic2014methodology, arghode2015experimental, raju2017heat, tomic2018perforated, husin2021modification} and flow-conditioning components \cite{laws1995further, spearman1996comparison, laribi2003comparative, xiong2003velocity, hoffmann2011effect, yaici20143d, laribi2015discharge}, in flame stabilization and control in combustion chambers \cite{noiray2007passive, oh2016stabilization, rashwan2017experimental, kim2020effects, younesian2021experimental, younesian2022visualization}, in turbulence manipulation \cite{naot1980penetration, liu2004generation, liu2007turbulent, dhineshkumar2013large}, and in the reduction of aerodynamic noise \cite{sakaliyski2007aero, rubio2019mechanisms, laffay2020experimental, sumesh2021aerodynamic, zamponi2025noise}. By redistributing momentum and promoting flow uniformity, perforated plates can significantly enhance the performance and reliability of downstream components. The flow through a perforated plate is typically characterized by flow separation at the pore inlet, acceleration and contraction toward a vena contracta, followed by flow expansion downstream of the vena contracta. The dominant contribution to pressure loss arises during this expansion process, where irreversible energy dissipation occurs due to flow separation and mixing \cite{miller1990internal}. For a visual illustration of this underlying physical mechanism, readers are referred to the eleventh figure in Bayazit et al.~\cite{bayazit2014perforated}, which provides a clear depiction of the flow behavior of the flow through a perforated plate.

From a design perspective, perforated plates are often required to provide a prescribed pressure loss to ensure adequate flow uniformity or turbulence characteristics. For example, in heat transfer applications, flow maldistribution is a major challenge that degrades the performance of heat transfer devices \cite{yaici20143d, pacio2010study, nielsen2013influence, beckedorff2022flow}. In practice, perforated plates are commonly employed as flow-conditioning elements to promote a more uniform velocity distribution upstream of heat transfer surfaces. Achieving effective flow uniformity requires the perforated plates to introduce a sufficiently large pressure drop. Consequently, accurate prediction of the pressure loss across a perforated plate is essential for optimal design. Overestimating pressure losses may lead to overly conservative designs and unnecessary energy penalties, whereas underestimating them can result in inadequate flow control and degraded thermal performance. In addition to improving flow uniformity, perforated plates can enhance turbulence homogeneity, thereby further improving overall system performance. However, their use also entails an increase in aerodynamic drag, leading to higher energy consumption. It is therefore of significant practical importance to quantify pressure losses across perforated plates and to develop reliable predictive models that enable a priori estimation of these losses, allowing an optimal compromise between flow conditioning effectiveness and associated drag penalties. 

\begin{figure}
\centering
\begin{overpic}[width=0.3\textwidth,trim=4.5cm -2cm 0cm 0cmm, clip]{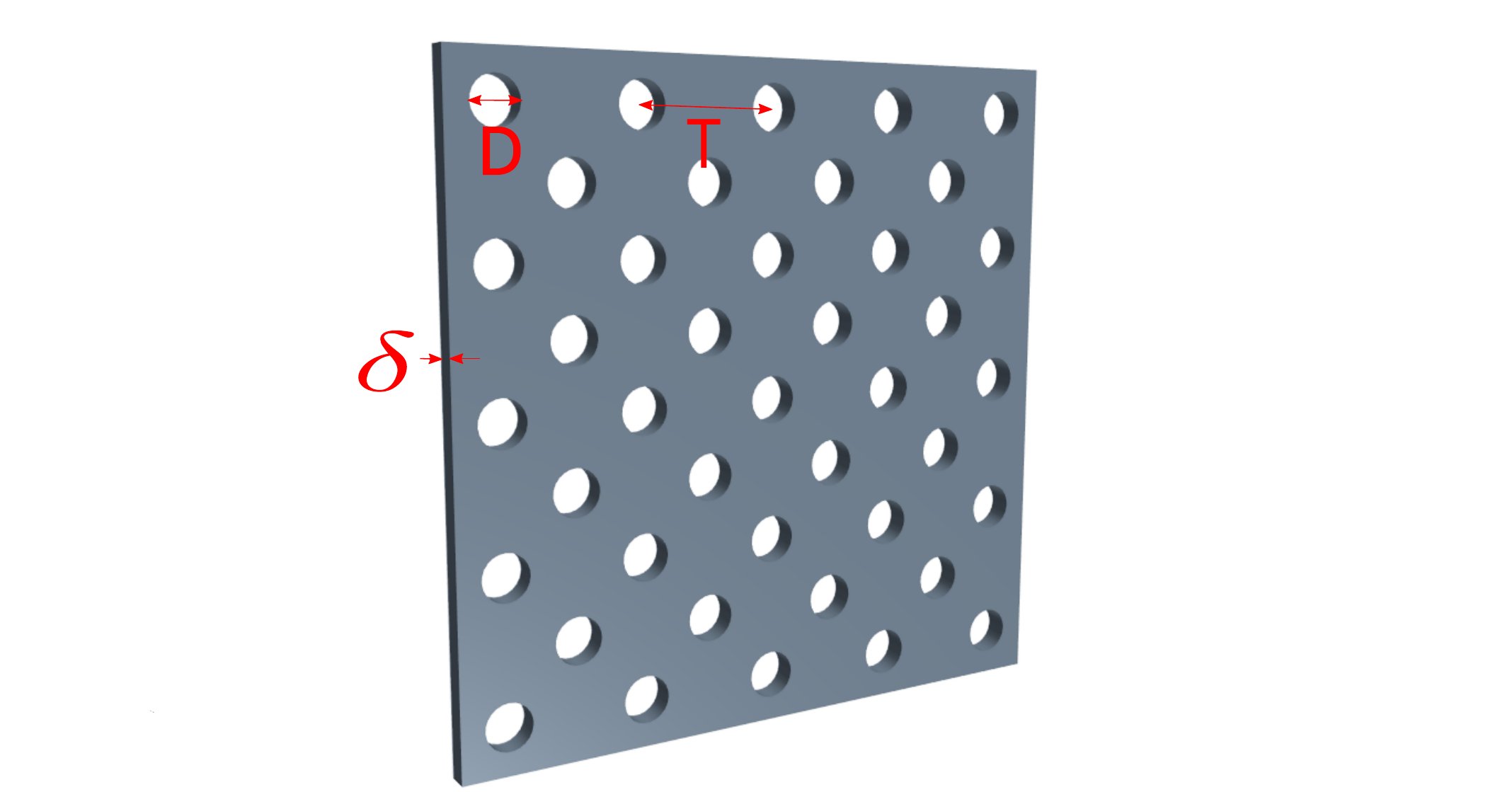} \put(5,68){($a$)} \end{overpic} 
\begin{overpic}[width=0.45\textwidth,trim=0cm 3.5cm 0cm -2cm, clip]{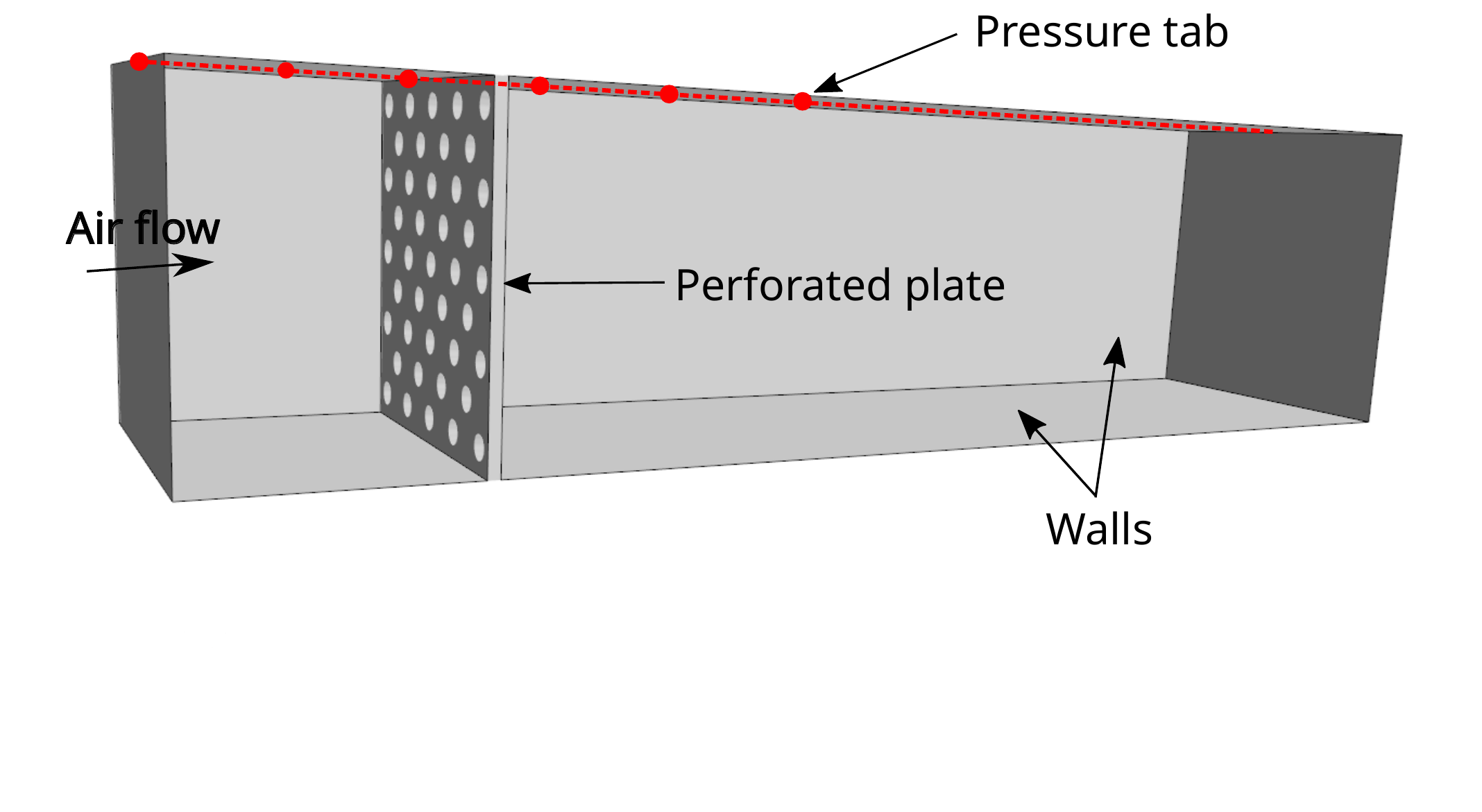} \put(0,45){($b$)} \end{overpic} 
\\
\caption{Illustrations of (a) a perforated plate with circular perforations and (b) the B2A wind-tunnel test section.}
\label{fig: sketch_perforated plate_B2A}
\end{figure}

Pressure losses across perforated plates usually depend on several geometric parameters of perforated plates, such as plate thickness, pore diameter, and pore spacing, as well as on flow conditions and regimes. Capturing the combined effects of these parameters in a predictive manner remains a challenging task. Previous research on pressure losses across perforated plates can be broadly classified into experimental and numerical studies. Yavuzkurt \& Catchen \cite{yavuzkurt2003dependence} experimentally investigated the dependence of pressure loss on air speed using wind-tunnel tests conducted on seven perforated plates with varying thicknesses and porosities. Measurements were performed at air speeds of up to $76$ m/s. For most plates, the pressure loss was found to scale linearly with the square of the air speed, indicating that inertial effects dominate the pressure loss mechanism. However, for a specific plate configuration, slight deviations from this linear behavior were observed, attributed to wake dynamics characterized by large-scale flow oscillations downstream of the plate. Recently, Méry \& Sebbane \cite{mery2023aerodynamic} experimentally examined the pressure-loss characteristics of perforated plates with circular pores, considering a range of pore diameters, spacings, and plate thicknesses, within the framework of the European Union H2020 INVENTOR project. Figure \ref{fig: sketch_perforated plate_B2A}(a) illustrates a perforated plate of thickness $\delta$, with pores of diameter $D$ separated by a spacing $T$. Variations in pore diameter and spacing lead to different plate porosities, $\varepsilon$, defined as the ratio of the total pore area to the total plate area. Their findings demonstrated that the pressure loss coefficient, defined as the pressure drop normalized by the dynamic pressure, is largely independent of flow velocity but strongly dependent on the relative plate thickness $\delta/D$. Using numerical simulations, Bayazit et al. \cite{bayazit2014perforated} studied pressure losses across perforated plates with two relative thicknesses ($\delta/D=0.5$ and $1.0$) and three porosities ($\varepsilon=0.2$, $0.35$ and $0.5$) over a wide range of Reynolds numbers spanning laminar to turbulent regimes. For laminar flows, thicker plates and lower porosities resulted in higher pressure losses. In contrast, under turbulent conditions, thinner plates produced larger pressure drops, while reduced porosity continued to increase the pressure loss. These contrasting trends were attributed to differences in flow separation and reattachment mechanisms associated with plate thickness across different flow regimes. Bae \& Kim \cite{bae2016numerical} numerically studied the pressure losses of laminar flows through thick perforated plates ($\delta/D \geq 1$) at pore-level Reynolds numbers up to 25 and proposed a simple predictive model for pressure losses in laminar regimes. While their model showed good agreement with experimental data, its applicability is restricted to laminar flows. In addition, several empirical models for pressure loss across perforated plates have been proposed, including those by Idelchik \cite{idelchik1986handbook,idelchik1994handbook}, Miller \cite{miller1990internal}, Kast \cite{kast2010pressure}, Holt et al. \cite{holt2011cavitation}, ESDU \cite{engineering1981flow}, Malavasi et al. \cite{malavasi2012pressure}, and Li et al. \cite{li2024pressure}, etc. Some of these empirical models are provided in the Appendix for reference. Although these models can generally be applied to turbulent flows through perforated plates with different hole arrangements, such as staggered or in-line configurations \cite{bayazit2014perforated,tanner2019flow}, they are very complicated including parameters to determine using empirical formulae. The purpose of the present study is to develop data-driven models for accurately predicting pressure losses across perforated plates in turbulent flow regimes, providing a simpler and more general alternative to existing empirical models.

In this paper, two data-driven models, a Kriging model and a NN model, are proposed to predict pressure losses across perforated plates using the experimental data of Yavuzkurt \& Catchen \cite{yavuzkurt2003dependence} and Méry \& Sebbane \cite{mery2023aerodynamic}. Compared with alternative approaches, such as simple loss coefficients, empirical correlations, detailed CFD simulations, or experimentally measured resistance curves, data-driven models provide a favorable balance between predictive accuracy and cost. Simple loss coefficients typically account for parameters such as the opening ratio or porosity of perforated plates while neglecting the influence of plate thickness. High-fidelity CFD simulations can accurately resolve the flow through individual pores but are computationally expensive because all pore-scale flow structures must be captured. Although RANS simulations are more affordable, they generally cannot accurately predict complex pore-scale flow phenomena, including flow separation at the pore leading edge and the formation of closed or open recirculation zones within the pores. Experimental measurements can provide highly accurate resistance characteristics. However, they are costly and time-consuming to obtain, and the amount of available experimental data remains limited. In this study, the proposed data-driven models are validated against a test dataset and are compared with several previously published models \cite{bae2016numerical, idelchik1994handbook, kast2010pressure, miller1990internal, holt2011cavitation}. Furthermore, to demonstrate the applicability of the new models in numerical simulations, two-dimensional channel-flow simulations are performed using RANS with the new models added as pressure-drop source terms to the momentum equations. The RANS-predicted pressure losses are subsequently compared with the model-predicted pressure losses.

The remainder of this paper is organized as follows. Section \ref{methodology} describes the methodology, including the experimental data used for training and validating the proposed data-driven models, as well as the formulation of the Kriging and NN models. Section \ref{kriging_NN_predictions} presents the pressure-loss predictions of the Kriging and NN models and compares them with previously published models. Section \ref{application_model} then demonstrates the application of the proposed models in numerical simulations. Finally, concluding remarks are provided in Section \ref{conclusion}. A discussion of existing pressure-loss models for perforated plates is included in the Appendix.

\section{Methodology}\label{methodology}

\subsection{Description of the experimental data}\label{experimental_data}

In the present study, experimental data reported by Yavuzkurt \& Catchen \cite{yavuzkurt2003dependence} and by Méry \& Sebbane \cite{mery2023aerodynamic} are used to train and validate the proposed data-driven models for predicting pressure losses across perforated plates. The measurements were conducted at the Pennsylvania State University and at the Aero-thermo-Acoustics Bench (B2A) of the Office National d'Etudes et de Recherches Aérospatiales (ONERA), respectively. 

The measurements of Yavuzkurt \& Catchen \cite{yavuzkurt2003dependence} were performed in a pipe-flow apparatus. The pipe has an internal diameter of $90.17$ mm. Airflow was supplied by a 5 hp blower operating in suction mode, with the flow rate regulated using a bypass line and an inline valve. To minimize the influence of local turbulence generated near the perforated plate, pressure taps were positioned $710$ mm upstream and downstream of the plate. The airflow velocity was measured using an anemometer placed near the upstream opening of the entrance pipe, where the velocity distribution across the inlet area was relatively uniform. The pressure losses were measured over a range from $0$ to $3500$ Pa. Additional details regarding the experimental setup are provided in Ref.~\cite{yavuzkurt2003dependence}.

The measurements of Méry \& Sebbane \cite{mery2023aerodynamic} were carried out in the B2A wind tunnel, which features a 0.2 m long test section with a square cross section of 50 mm $\times$ 50 mm. A schematic of the wind-tunnel test section is shown in Fig. \ref{fig: sketch_perforated plate_B2A}(b). The pressure drop across the perforated plate was measured using static pressure taps located on the top wall of the test section, both upstream and downstream of the plate. The positions of the pressure taps are listed in Table~\ref{tab:pressure_taps}, with PS1 taken as the reference \cite{mery2023aerodynamic}. The perforated plates were installed midway between PS3 and PS4, corresponding to a location at 37.5 mm downstream of the PS1 pressure tap. During the experiments, the airflow velocities at the wind-tunnel entrance were set to $U_{0} = 16.6$,  $25$, and $35$ m/s. These measurements were recently conducted within the framework of the European Union H2020 INVENTOR project. Further details regarding the B2A wind tunnel and the experimental procedures can be found in Refs.~\cite{minotti2008characterization, mery2023aerodynamic}.

\begin{table}[width=.65\linewidth,cols=4,pos=h]
\begin{threeparttable}
\caption{Streamwise positions of the pressure taps.}\label{tab:pressure_taps}
\begin{tabular*}{\tblwidth}{@{} cccccccc@{} }
\toprule
      Pressure taps  & PS1 & PS2 & PS3 & PS4 & PS5 & PS6 & PS7 \\[3pt]
\midrule
       Streamwise position (mm)   & 0 & 15 & 30 & 45 & 60 & 75 & 215\\
\bottomrule
\end{tabular*}
\end{threeparttable}
\end{table}

Table \ref{tab:Yuvuzkurt_Catchen_Mery_Sebbane} summarizes the pore diameter $D$, pore spacing $T$, plate thickness $\delta$, and porosity $\varepsilon$ of the perforated plates used in the experiments conducted by Yavuzkurt \& Catchen \cite{yavuzkurt2003dependence} and by Méry \& Sebbane \cite{mery2023aerodynamic}. The porosity, defined as the ratio of the total pore area to the overall area of the perforated plate, can be calculated from the pore diameter and pore spacing. In this study, 70\% of the available experimental data (Plates 1–8) are used to train the models, while the remaining 30\% (Plates 9–11) are reserved as a test dataset. Given the scarcity of experimental data on pressure losses across perforated plates in the literature and the limited total of only 11 data points in this study, this data partitioning represents a pragmatic compromise between ensuring sufficient information for effective model training and maintaining an independent dataset for an unbiased assessment of predictive performance.

\begin{table}[width=.48\linewidth,cols=4,pos=h]
\begin{threeparttable}
\caption{Parameters of the perforated plates used in the experiments of Yavuzkurt \& Catchen \cite{yavuzkurt2003dependence} (Plates 1-7) and Méry \& Sebbane \cite{mery2023aerodynamic} (Plates 8-11).}\label{tab:Yuvuzkurt_Catchen_Mery_Sebbane}
\begin{tabular*}{\tblwidth}{@{} cccccc@{} }
\toprule
      Plate  & $D$ (mm) & $T$ (mm) & $\delta$ (mm) & $\varepsilon$ & $\delta/D$ \\[3pt]
\midrule
       1   & 1.60 & 3.18 & 1.02 & 0.227 & 0.638 \\
       2   & 1.91 & 2.54 & 0.81 & 0.510 & 0.424 \\       
       3   & 2.38 & 6.88 & 0.91 & 0.109 & 0.382 \\
       4   & 2.78 & 4.76 & 0.81 & 0.309 & 0.291 \\
       5   & 3.18 & 4.76 & 0.91 & 0.403 & 0.286 \\       
       6   & 4.76 & 9.53 & 0.97 & 0.227 & 0.204 \\
       7   & 3.18 & 4.76 & 3.18 & 0.403 & 1.000 \\
       8   & 5.00 & 6.00 & 1.00 & 0.623 & 0.200 \\
       9   & 4.00 & 6.00 & 1.00 & 0.403 & 0.250 \\       
       10   & 2.00 & 3.00 & 2.00 & 0.403 & 1.000 \\
       11   & 2.00 & 3.00 & 1.00 & 0.403 & 0.500 \\
\bottomrule
\end{tabular*}
\end{threeparttable}
\end{table}

\subsection{Kriging model}
In this study, we want to predict the pressure loss of perforated plates at any given plate porosity $\epsilon$ and thickness ratio $\delta/D$. Without loss of generality, assume that we wish to make a prediction at a given point $\mathbf{x}$. $Y(\mathbf{x})$ is a realization of a random variable that is normally distributed with mean $\mu$ and variance $\sigma^2$. Now consider two points $\mathbf{x}_i$ and $\mathbf{x}_j$. The correlation between the random variables is given by~\cite{jones2001taxonomy}
\begin {equation}\label{correlation}
\text{Corr}\bigl(Y(\mathbf{x}_i),\, Y(\mathbf{x}_j)\bigr)
= \exp\!\left(
    - \sum_{\ell=1}^d 
      \theta_\ell \, \lvert x_{i\ell} - x_{j\ell} \rvert^{p_\ell}
  \right),
\end {equation}
where $\theta_\ell$ controls how quickly the correlation decreases as one moves along the $\ell$th coordinate direction. The parameter $p_\ell$ governs the smoothness of the function in that direction: values of $p_\ell$ close to 2 correspond to smoother functions, whereas values near 0 capture rough, non-differentiable behavior. For $n$ points, the covariance function is given by~\cite{jones2001taxonomy}
\begin {equation}\label{covariance}
\operatorname{cov}\!\left[ Y\!\left(\mathbf{x}_i\right),\, Y\!\left(\mathbf{x}_j\right) \right]
= \sigma^{2} R,
\end {equation}
where $R$ is a $n \times n$ matrix with the $(i, j)$ element given by Eq. (\ref{correlation}). The values of $\mu$, $\sigma^2$, $\theta_\ell$, and $p_\ell$ ($\ell = 1, \ldots, d$) are then estimated by maximizing the likelihood of the observed data. The Kriging predictor is given by~\cite{jones2001taxonomy}
\begin {equation}\label{kringing_predictor}
\hat{y}(\mathbf{x}^*) = \hat{\mu} + \mathbf{r}^T \mathbf{R}^{-1} ( \mathbf{y} - \mathbf{I}\hat{\mu} ),
\end {equation}
where $\mathbf{I}$ is a $n\times1$ vector of ones, and $\mathbf{r}$ denotes the vector of correlations of $Y(\mathbf{x}^*)$ with $Y(\mathbf{x}_i)$, for $i = 1, \ldots, n$:
\begin {equation}\label{corr_obs_unknown}
\mathbf{r} =
\begin{pmatrix}
\operatorname{Corr}\!\left( Y(\mathbf{x}^*),\, Y(\mathbf{x}_1) \right) \\
\vdots \\
\operatorname{Corr}\!\left( Y(\mathbf{x}^*),\, Y(\mathbf{x}_n) \right)
\end{pmatrix}.
\end {equation}

The mean-squared error of the Kriging predictor is~\cite{jones2001taxonomy}
\begin {equation}\label{mean_squared_error}
s^{2}(\mathbf{x}^*) = \hat{\sigma}^2 \left[\, 1 - \mathbf{r}^T\mathbf{R}^{-1}\mathbf{r}
\;+\; \frac{\left(1 - \mathbf{r}^T\mathbf{R}^{-1}\mathbf{r}\right)^2}{\mathbf{I}^T\mathbf{R}^{-1}\mathbf{I}} \right].
 \end {equation}
 
For convenience, one often uses the square root of the mean squared error, $s=\sqrt{s^{2}(\mathbf{x}^*)}$. This root mean squared error, also referred to as the “standard error”, provides a natural measure of uncertainty in the predictions.
 
 \subsection{Neural Network model}
In this work, a multilayer perceptron (MLP), one of the earliest and most widely adopted NN architectures, is employed. A simple MLP is adopted because the relationship between the two input variables and the pressure loss is a low-dimensional nonlinear regression problem.  A relatively standard MLP can give accurate predictions with minor prediction errors~\cite{fadaei2018application}. In this study, the MLP architecture comprises an input layer with two variables, a hidden layer, and an output layer with a single variable. Besides the NN architecture, the selection of input parameters is shown to have an important impact on the model prediction accuracy~\cite{sajjad2025physics}. Here, $\epsilon$ and $\delta/D$ are selected as input parameters. The input vector is then successively transformed through linear operations followed by the nonlinear Tanh activation function within the hidden layer, ultimately producing the normalized pressure drop $\Delta p/(\rho U_0^2)$ as the model output. The Tanh activation function is employed because it provides a smooth nonlinear mapping for the regression task. The fully connected MLP is implemented using an available framework~\cite{berguin2024jacobianenhanced}. 

To evaluate the predictive performance and generalizability of the NN model, the Leave-One-Out Cross Validation (LOOCV) is employed during training. Given the constraints of the training data size ($n=8$), traditional hold-out methods or $k$-fold cross-validation risk under-utilizing available data for training. LOOCV addresses this by iteratively training the model on $n-1$ training data points and validating it against the single excluded data point. This approach ensures that the model is trained on the maximum possible amount of data in each iteration. In this study, the LOOCV uses Bayesian optimization in Optuna to tune the hyperparameters. Neural networks with hyperparameters tuned via Bayesian optimization exhibit excellent predictive performance~\cite{rashidi2022applications}\color{black}. Owing to the limited size of the available dataset, the network architecture is intentionally kept simple to reduce the risk of overfitting. Accordingly, a single hidden layer with 2–4 neurons was considered. The learning rate $\alpha$ ranges from $10^{-4}$ to $10^{-1}$. The L2 regularization parameter $\lambda$ for mitigating overfitting ranges from $10^{-4}$ to $10^{-1}$. The NN surrogate is trained using full-batch Adam optimization. The Adam optimizer is used due to its robust and efficient optimization performance. The number of optimizer iterations per batch ranges from 300 to 1000. During training, a backtracking line search is employed to ensure stable optimization. Through the Bayesian optimization in Optuna, a neural network with 3 neurons, 768 optimizer iterations, and $\lambda=6.33\times10^{-4}$ was selected for the training and testing stages.

\section{Results}\label{results}

\subsection{Predictions of the Kriging and Neural Network models}\label{kriging_NN_predictions}

\begin{figure}
\centerline{\includegraphics[angle=0,width=12.0cm]
  {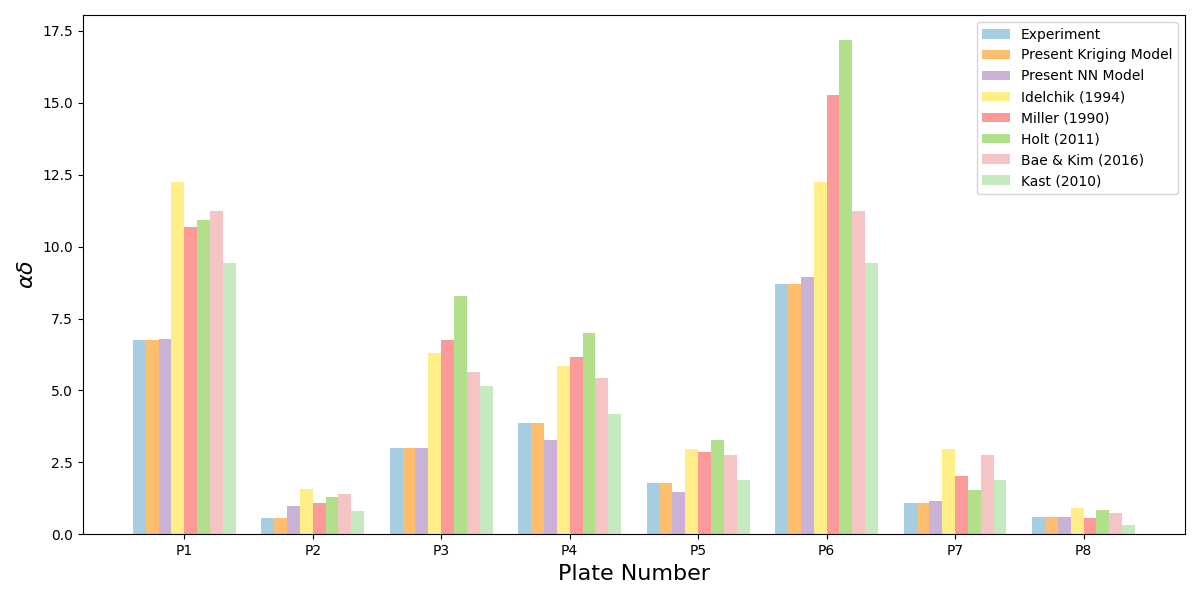}}
\caption{Comparison of the pressure drop predicted by the present Kriging and NN models with experimental pressure-drop data from the training dataset and with predictions from previously published models \cite{bae2016numerical, idelchik1994handbook, kast2010pressure, miller1990internal, holt2011cavitation}. For Plate 3, the results have been scaled by a factor of 0.1 to improve plot readability. Experimental pressure-drop values represent averages over all measurements corresponding to different pore-scale Reynolds numbers for each plate.} 
\label{fig:dp_training_dataset}
\end{figure}

\begin{figure}
\centerline{\includegraphics[angle=0,width=12.0cm]
  {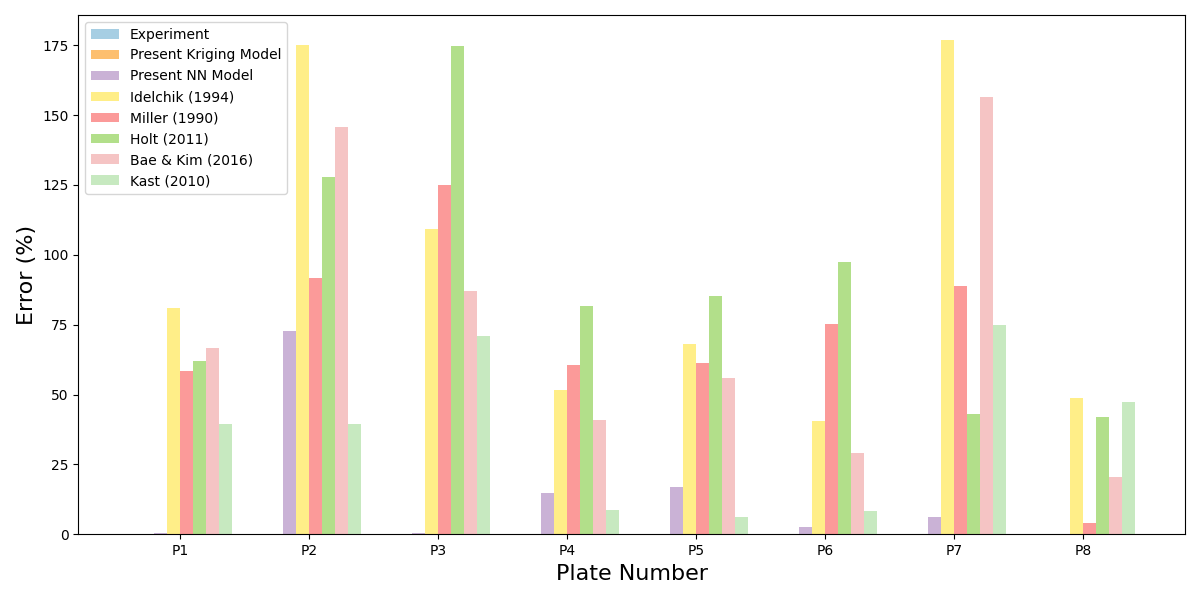}}
\caption{Comparison of prediction errors of different pressure-drop models relative to experimental data in the training dataset.} 
\label{fig:Error_training_dataset}
\end{figure}

Figure~\ref{fig:dp_training_dataset} compares the pressure drops predicted by the present Kriging and NN models with experimental pressure-drop data from the training dataset, as well as with predictions from previously published models \cite{bae2016numerical, idelchik1994handbook, kast2010pressure, miller1990internal, holt2011cavitation}. The dimensionless pressure drop, as mathematically shown in the Appendix, consists of a Darcy term and a Forchheimer term ($\alpha \delta$). Under the turbulent flow conditions considered in the present study, the Forchheimer contribution is significantly larger than the Darcy contribution. Consequently, the dimensionless pressure drop is dominated by the Forchheimer term and can be approximated by $\alpha \delta$. For the Bae \& Kim model \cite{bae2016numerical}, only the Forchheimer component of the pressure drop is considered. As shown in Fig.~\ref{fig:dp_training_dataset}, both the Kriging and NN models generally provide more accurate predictions of the training data than the previously published models. Moreover, the pressure-drop predictions obtained from the Kriging model show better agreement with the experimental training data than those produced by the NN model. This conclusion is further supported by a comparison of the prediction errors of the different pressure-drop models relative to the experimental data, as presented in Fig. \ref{fig:Error_training_dataset}. The Kriging model exhibits no noticeable errors, while the NN model shows only some deviations for most of the plates. As shown in Table~\ref{tab:error_reductions}, the prediction errors for the eight perforated plates are 0.62\%, 76.92\%, 0, 15.22\%, 19.04\%, 2.9\%, 8.0\%, and 0, respectively. While the error for one case reaches 76.92\%, it remains substantially lower than the errors produced by several existing models, which exceed 100\% for the same perforated plates. The error reductions
achieved by the Kriging and NN models are presented in Table~\ref{tab:error_reductions}. Overall, the proposed Kriging model provides improved prediction accuracy compared with existing models. In addition, the Root-Mean-Square (RMS) errors of the Kriging and NN models evaluated on the training and test datasets are presented in Table~\ref{tab:RMS}. Nevertheless, it should be noted that comparisons based solely on the training data are not fully sufficient to validate the newly developed models, as they were trained using the same dataset. A more convincing assessment of their predictive capability requires evaluating the models against an independent test dataset.

\renewcommand{\figurename}{Fig.}

\renewcommand{\figurename}{Fig.}

\begin{table}[width=1.00\linewidth,cols=4,pos=h]
\begin{threeparttable}
\caption{Prediction errors (in percentage) of the Kriging and NN models relative to experimental data, and the corresponding error reduction compared with existing models. The first and second numbers in parentheses indicate the error reductions achieved by the Kriging and NN models, respectively.}\label{tab:error_reductions}
\begin{tabular*}{\tblwidth}{@{} cccccccc@{} }
\toprule
      Plate  & Kriging & NN & Idelchik (1994) & Miller (1990) & Holt (2011) & Bae\&Kim (2016) & Kast (2010) \\[3pt]
\midrule
       1   & 0.00 & 0.62 & (81.37, 80.75) & (57.76, 57.14) & (61.49, 60.87) & (66.46, 65.84) & (39.13, 38.51) \\
       2   & 0.00 & 76.92 & (184.61, 107.69) & (92.31, 15.38) & (130.77, 53.85) & (153.85, 76.92) & (46.15, -30.77) \\       
       3   & 0.00 & 0.00 & (111.27, 111.27) & (126.76, 126.76) & (177.46, 177.46) & (88.73, 88.73) & (71.83, 71.83) \\
       4   & 0.00 & 15.22 & (51.09, 35.87) & (59.78, 44.57) & (80.43, 65.22) & (40.22, 25.00) & (7.61, -7.61) \\
       5   & 0.00 & 19.04 & (66.67, 47.62) & (61.90, 42.86) & (83.33, 64.29) & (54.76, 35.71) & (4.76, -14.28) \\       
       6   & 0.00 & 2.90 & (41.06, 38.16) & (75.85, 72.95) & (97.58, 94.68) & (29.47, 26.57) & (8.70, 5.80) \\
       7   & 0.00 & 8.00 & (180.00, 172.00) & (92.00, 84.00) & (44.00, 36.00) & (160.00, 152.00) & (76.00, 68.00) \\
       8   & 0.00 & 0.00 & (50.00, 50.00) & (7.14, 7.14) & (42.86, 42.86) & (21.43, 21.43) & (50.00, 50.00) \\
       9   & 10.66 & 9.13 & (70.56, 72.08) & (71.07, 72.59) & (97.97, 99.49)  & (57.87, 59.39) & (4.06, 5.58) \\       
       10   & 21.34 & 15.85 & (96.95, 102.44) & (27.44, 32.93) & (-8.54, -3.05) & (81.10, 86.59) & (16.46, 21.95) \\
       11   & 40.26 & 47.40 & (-24.35, -31.49) & (-32.47, -39.61) & (-39.61, -46.75) & (-32.47, -39.61) & (-13.64, -20.78) \\
\bottomrule
\end{tabular*}
\end{threeparttable}
\end{table}

\begin{table}[width=.23\linewidth,cols=4,pos=h]
\begin{threeparttable}
\caption{RMS errors of the Kriging and NN models evaluated on the training and test datasets.}\label{tab:RMS}
\begin{tabular*}{\tblwidth}{@{}ccc@{} }
\toprule
      Dataset  & Kriging & NN \\[3pt]
\midrule
       Training  & 0.000 & 0.297 \\
       Test   & 0.628 & 0.719 \\       
\bottomrule
\end{tabular*}
\end{threeparttable}
\end{table}

\renewcommand{\figurename}{Fig.}
\begin{figure}
\centerline{\includegraphics[angle=0,width=12.0cm]
  {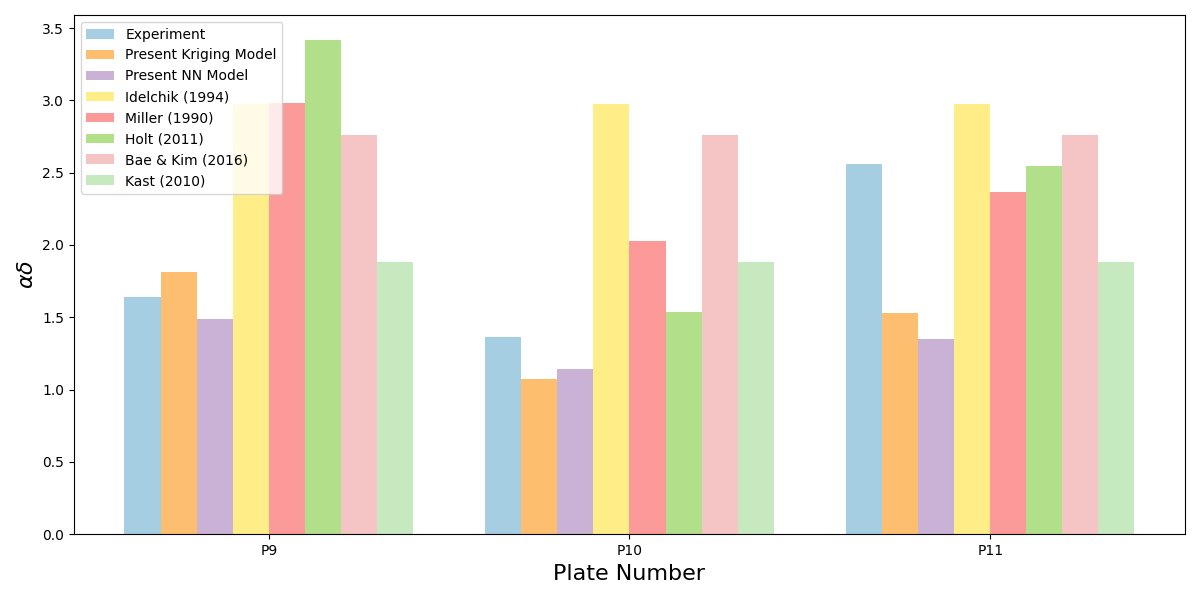}}
\caption{Comparison of the pressure drop predicted by the present Kriging and NN models with experimental pressure-drop data from the test dataset and with predictions from previously published models \cite{bae2016numerical, idelchik1994handbook, kast2010pressure, miller1990internal, holt2011cavitation}. Experimental pressure-drop values represent averages over all measurements corresponding to different pore-scale Reynolds numbers for each plate.} 
\label{fig:dp_test_dataset}
\end{figure}

\renewcommand{\figurename}{Fig.}
\begin{figure}
\centerline{\includegraphics[angle=0,width=12.0cm]
  {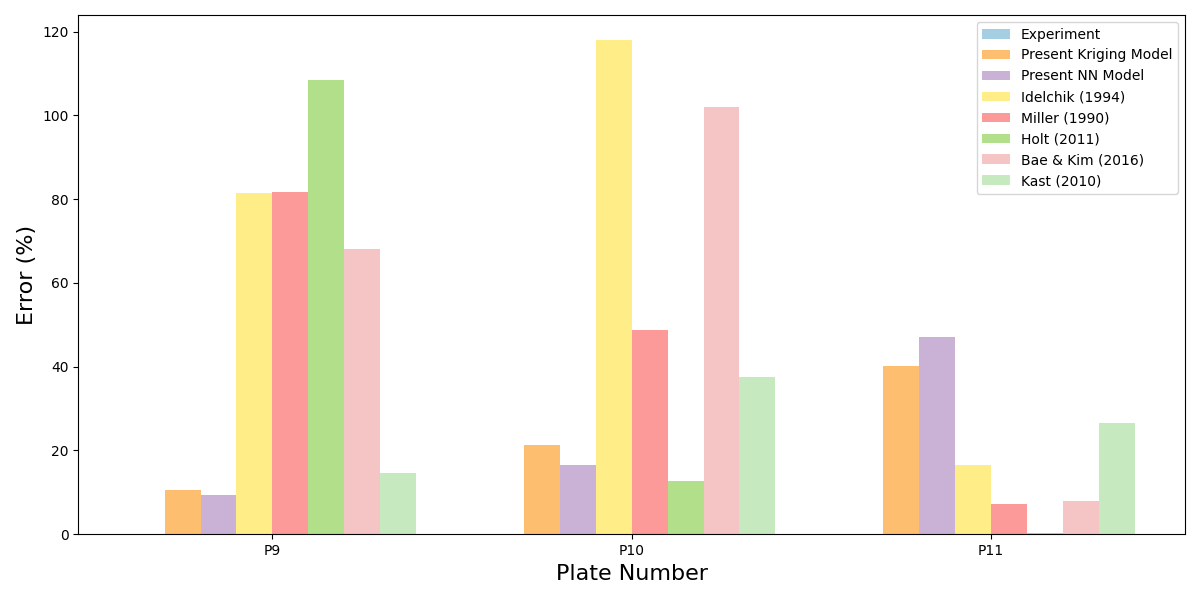}}
\caption{Comparison of prediction errors of different pressure-drop models relative to experimental data in the test dataset.} 
\label{fig:Error_test_dataset}
\end{figure}

Figure \ref{fig:dp_test_dataset} compares the pressure drops predicted by the present Kriging and NN models with experimental pressure-drop data from the test dataset, as well as with predictions from previously published models \cite{bae2016numerical, idelchik1994handbook, kast2010pressure, miller1990internal, holt2011cavitation}. When evaluated against the test data, both the Kriging and NN models yield reasonable predictions with relatively small discrepancies for Plates 9 and 10, compared to previously published models. However, a noticeable discrepancy is observed for Plate 11. As shown in Table~\ref{tab:error_reductions}, for the three perforated plates, the prediction errors of the Kriging model are 10.66\%, 21.34\%, and 40.26\%, respectively, whereas the corresponding errors of the NN model are 9.13\%, 15.85\%, and 47.4\%, respectively. Although the prediction errors of the Kriging and NN models for Plate 11 are larger than those for Plates 9 and 10, they remain smaller than the errors produced by some other models for Plates 9 and 10, as illustrated in Figure~\ref{fig:Error_test_dataset} and Table~\ref{tab:error_reductions}. In fact, this discrepancy can be attributed to the limited amount of available experimental data used in model training. In the future, as more experimental data become available in the literature, this limitation can be readily alleviated. Furthermore, it is noteworthy that the prediction error of certain individual models (e.g., Holt \cite{holt2011cavitation} for Plate 9, and Idelchik \cite{idelchik1994handbook} and Bae \& Kim \cite{bae2016numerical} for Plate 10) can exceed 100\%. This highlights the considerable challenge of accurately modelling pressure losses in the complex problem of flow through perforated plates.

\begin{figure}
\centering
\begin{overpic}[width=0.49\textwidth,trim=0cm 0cm 0cm 0cmm, clip]{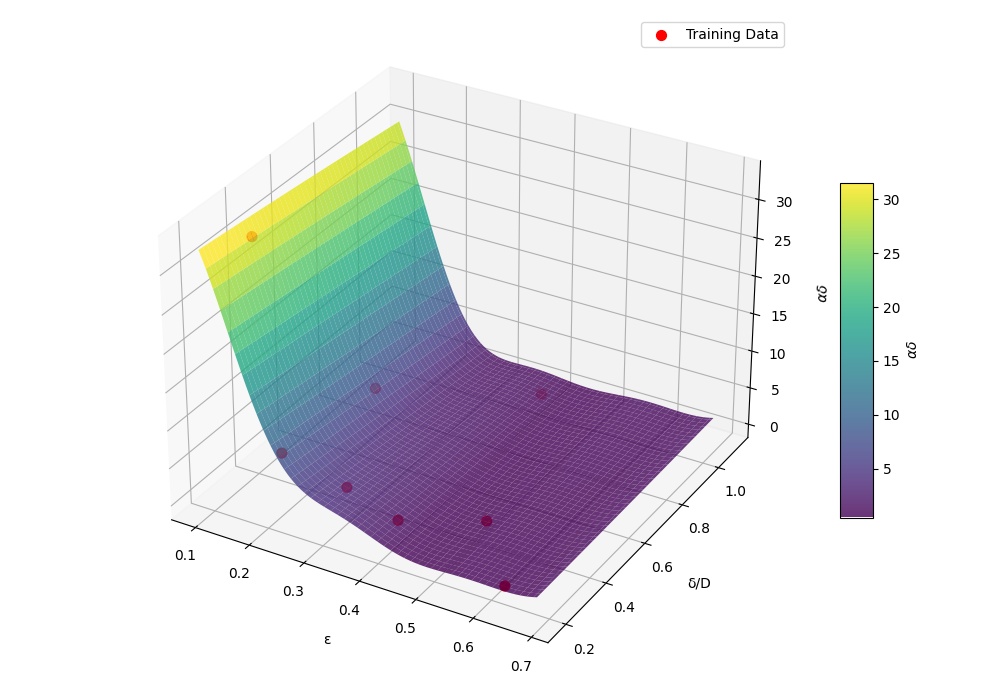} \put(10,65){($a$)} \end{overpic} 
\begin{overpic}[width=0.49\textwidth,trim=0cm 0cm 0cm 0cm, clip]{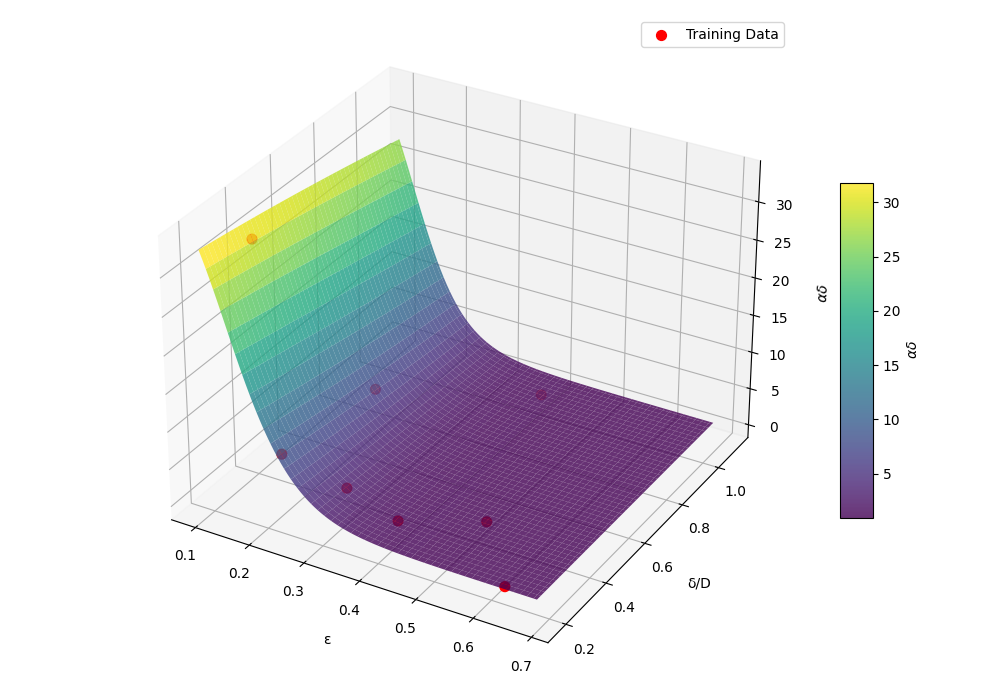} \put(10,65){($b$)} \end{overpic} 
\\
\caption{Response surfaces of the present (a) Kriging model and (b) NN model, shown alongside the training data points.}
\label{fig: response_surface}
\end{figure}

Figure~\ref{fig: response_surface} presents the response surfaces of the present Kriging and NN models, along with the training data points. For both models, the response surfaces closely follow the trends exhibited by the training data. The pressure drop is highly sensitive to porosity, particularly at low porosity values, where the response surfaces are markedly steeper. As porosity increases, this sensitivity gradually diminishes, leading to progressively flatter response surfaces. This behavior can be attributed to the pronounced blockage effect of perforated plates with low porosity, where small changes in open area produce substantial variations in flow resistance. In contrast, once the porosity exceeds a certain threshold, the blockage effect becomes negligible, and further increases in open area have only a minor influence on the pressure drop. In contrast, the pressure drop exhibits relatively weak sensitivity to the thickness ratio.

\renewcommand{\figurename}{Fig.}
\begin{figure}
\centerline{\includegraphics[angle=0,width=17cm,trim=2cm 2cm 0cm 2cmm, clip]
  {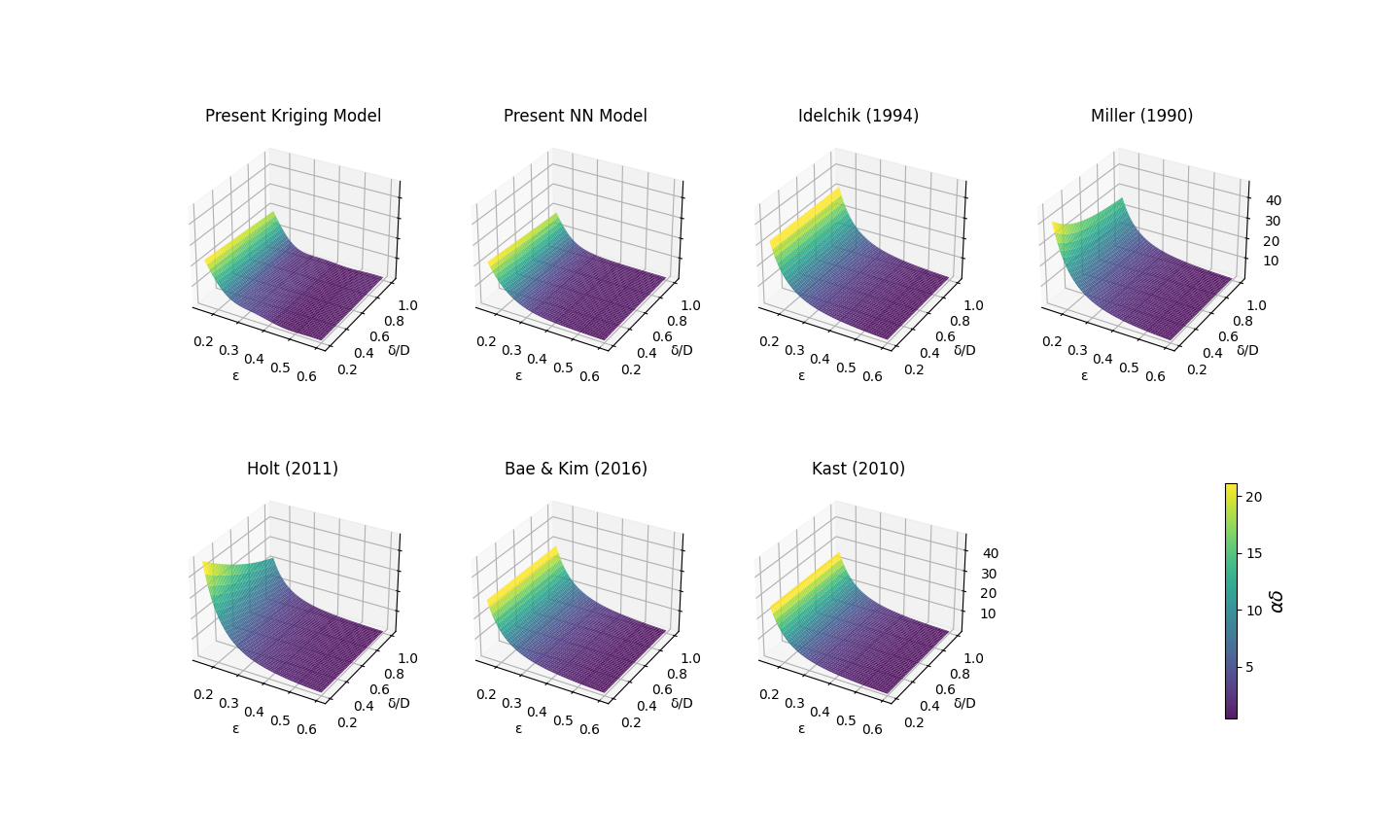}}
\caption{Comparison of response surfaces from the present Kriging and NN models and previous models \cite{bae2016numerical, idelchik1994handbook, kast2010pressure, miller1990internal, holt2011cavitation}.} 
\label{fig:comparison_response_surfaces}
\end{figure}

Figure~\ref{fig:comparison_response_surfaces} compares the response surfaces produced by the present Kriging and NN models with those from previously published models \cite{bae2016numerical, idelchik1994handbook, kast2010pressure, miller1990internal, holt2011cavitation}. As expected, all models predict decreasing pressure drops with increasing porosity. However, the response surfaces exhibit different slopes at low porosity among the models. This variation stems from the different dependence of the empirical correlations on porosity. Specifically, the empirical formulae in the Appendix show that porosity, $\epsilon$, appears with different exponents in the denominator of the correlations, causing the predicted pressure drop to increase at different rates as porosity decreases. Consequently, although all models predict high pressure drops at low porosities, their sensitivities differ. The influence of the thickness ratio is more pronounced in the models proposed by Miller \cite{miller1990internal} and Holt \cite{holt2011cavitation}. In particular, these models generally predict higher pressure drops at low thickness ratios than at high thickness ratios, with this trend becoming increasingly pronounced as porosity decreases. This behavior arises because the empirical correlations proposed by Miller \cite{miller1990internal} and Holt \cite{holt2011cavitation} depend not only on porosity but also on the thickness ratio, allowing variations in thickness ratio to influence the predicted pressure drop, especially at low porosity.

\renewcommand{\figurename}{Fig.}
\begin{figure}
\centerline{\includegraphics[angle=0,width=17cm,trim=-2cm 0cm 0cm 0cmm, clip]
  {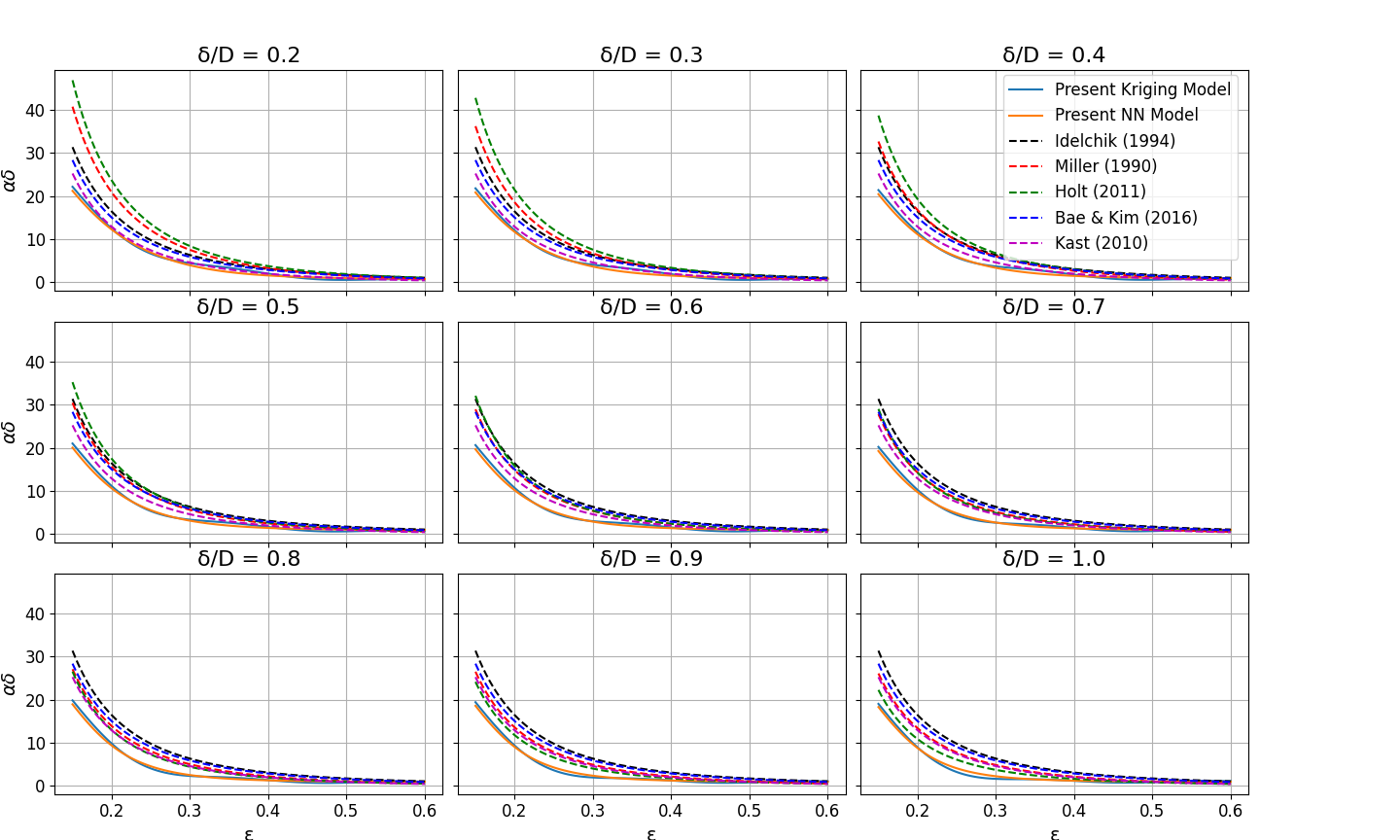}}
\caption{Comparison of pressure drops predicted by the present Kriging and NN models with those from previously published models \cite{bae2016numerical, idelchik1994handbook, kast2010pressure, miller1990internal, holt2011cavitation} for different thickness ratios $\delta/D$. } 
\label{fig:comparison_dp_at_different_thickness}
\end{figure}

Figure~\ref{fig:comparison_dp_at_different_thickness} compares the pressure drops predicted by the present Kriging and NN models with those obtained from previously published models \cite{bae2016numerical, idelchik1994handbook, kast2010pressure, miller1990internal, holt2011cavitation} for different thickness ratios $\delta/D$. At low porosity levels, the Kriging and NN models generally predict lower pressure drops than the previously published models. As discussed earlier, this discrepancy arises from the different dependence of the empirical correlations on porosity. In contrast, at high porosity levels, all models produce nearly identical pressure-drop predictions. This convergence occurs because, as the porosity $\epsilon$ approaches unity, $\epsilon$ raised to different exponents also approaches unity. Consequently, the influence of the exponent diminishes, and the predicted pressure drops converge to similar values. Overall, the agreement among the models is excellent at high porosity, whereas deviations emerge at low porosity levels.

\subsection{Integration of the present models in numerical simulations}\label{application_model}

\subsubsection{Numerical modelling of perforated plates in two-dimensional channel flows}

Although many CFD applications require additional physical effects to be modeled, such as flow straightening, momentum recovery, and discharge characteristics, the present models do not capture these flow phenomena often associated with compact representations of perforated plates. The present study focuses exclusively on flow resistances, i.e. pressure drops. Flow straightening suppresses swirl and promotes a more uniform downstream velocity profile. As the jets emerging from individual pores merge, part of the pressure is recovered through momentum exchange between adjacent jets. However, irreversible energy dissipation associated with jet mixing prevents complete momentum recovery. In addition, the flow discharging through a perforated plate follows the principles of orifice flow. In practical applications, for perforated plates with very fine pores, directly meshing the plate geometry is highly challenging, as it would generate an excessively large number of computational cells. In such cases, it is preferable to use a numerical model that represents the effects of flow through perforated plates. This approach is computationally efficient because it eliminates the need to explicitly mesh and resolve the individual pores of the perforated plate. Similar to the modelling of other porous media \cite{okolo2017numerical, zhu2020numerical, li2023numerical, li2024mitigation, li2024numerical}, the influence of perforated plates on the flow can be incorporated by introducing a pressure-drop source term into the momentum equations. Specifically, within the perforated-plate region, a volumetric source term $S$ is added to the right-hand side (RHS) of the momentum equations, resulting in the RANS momentum equations as formulated in \cite{ccmUG}
\begin {equation} \label{governing_eqn}
\frac{\partial (\rho \bar{u}_i)}{\partial t} + \frac{\partial (\rho \bar{u}_i\bar{u}_j)}{\partial x_j}  = - \frac{\partial \bar{p}}{\partial x_i} + \frac{\partial \bar{\sigma}_{ij}}{\partial x_j} + \frac{\partial \sigma_{ij, RANS}}{\partial x_j} + S_i,
\end {equation}
where $\bar{u}_i$ represents the velocity components (for two-dimensional flows, $i=$1, 2), $\bar{p}$ is the pressure, $\bar{\sigma}_{ij}$ is the viscous stress tensor, and $\sigma_{ij, RANS}$ denotes the Reynolds-stress tensor. Outside the perforated-plate region, the governing equations reduce to the standard RANS momentum equations, i.e., without the additional source term $S_i$ in Eq. (\ref{governing_eqn}). Across the entire flow domain, including the perforated-plate region, the continuity equation remains unchanged, since the law of mass conservation continues to hold within the porous region. When the two-dimensional airflow approaches the perforated plate at zero incidence angle, the source term is expressed as

\begin{equation}\label{source_term}
     \begin{bmatrix}
           S_{1} \\
           S_{2} 
      \end{bmatrix} =   \begin{bmatrix}
                                      -\Delta p/\delta \\
                                      0
                                 \end{bmatrix}.
\end{equation}
Since the incoming flow is perpendicular to the perforated plate, the second component of the source term, $S_2$, is zero. During the simulation, the trained Kriging or NN model is loaded to predict the pressure loss from the specified perforated-plate parameters (i.e., porosity and thickness ratio). The predicted pressure loss is then incorporated into the source term of the momentum equation. The purpose of this implementation is limited to demonstrating a numerically robust method for imposing a prescribed pressure loss within a CFD simulation.

\subsubsection{Computational meshes and simulation setup}

\begin{figure}
\centerline{\includegraphics[angle=0,width=14.0cm, trim=1cm 2.5cm 0cm 0cmm, clip]
  {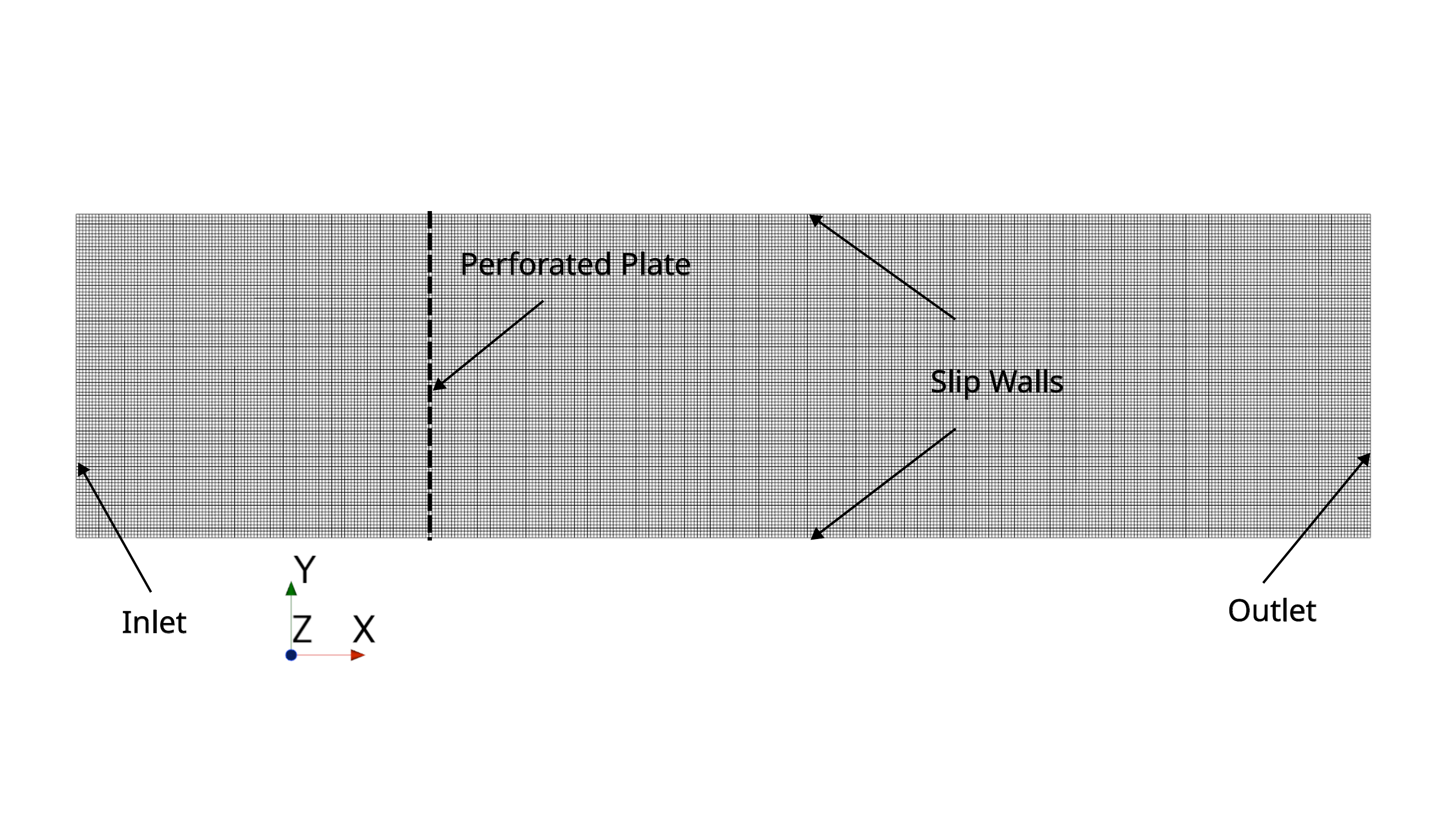}}
\caption{Two-dimensional coarse computational mesh with a uniform cell size of $\Delta=0.5$ mm used for the RANS simulations. The dashed line indicates the location at which the source term is applied in the momentum equations.} 
\label{fig:Mesh_2D}
\end{figure}

The computational domain is a rectangle measuring $200 \times 50 \: \mathrm{mm^2}$. Three mesh resolutions were considered to assess the sensitivity of the modeling approach to mesh resolution. In this study, three meshes, consisting of square cells with varying sizes, are employed in the simulations. The coarse mesh has a cell size of $\Delta = 0.5$ mm, while the medium and fine meshes have cell sizes of $\Delta = 0.25$ mm and $0.125$ mm, respectively. The total number of cells for the coarse, medium, and fine meshes is 40k, 160k, and 640k, respectively. Figure \ref{fig:Mesh_2D} presents the coarse computational mesh with boundary conditions. The perforated plate itself is not explicitly meshed. Instead, its effect is modelled by adding a momentum source term (i.e. Eq. (\ref{source_term})) at the location of the dashed line ($x=0$). Specifically, unless stated otherwise, for Plates 1 and 9, which have a thickness of 1 mm, the momentum source is applied to cells within $-0.5$ mm $<x<$ $0.5$ mm.

Steady RANS simulations are performed using the SST $k\text{--}\omega$ turbulence model within the commercial software Siemens STAR-CCM+ version 2022.1 \cite{ccm0}. It is also worth noting that, although the SST $k\text{--}\omega$ turbulence model is employed in the present study, the proposed Kriging and NN models are not restricted to this turbulence model. They can be readily coupled with other turbulence models, such as the standard or realizable $k\text{--}\varepsilon$ models and the Spalart–Allmaras model. The inlet velocity is set to $U_0 = 16.6$ m/s, corresponding to a Reynolds number of $Re=27,370$, based on the inlet velocity and the channel half-height. At the inlet boundary, the turbulence intensity is 6.2\%, and the turbulent viscosity ratio is 10, which determines the dissipation rate $\omega$.

\renewcommand{\figurename}{Fig.}

\subsubsection{RANS simulations using the present models as source terms}

\begin{figure}
\centerline{\includegraphics[angle=0,width=15.0cm]
  {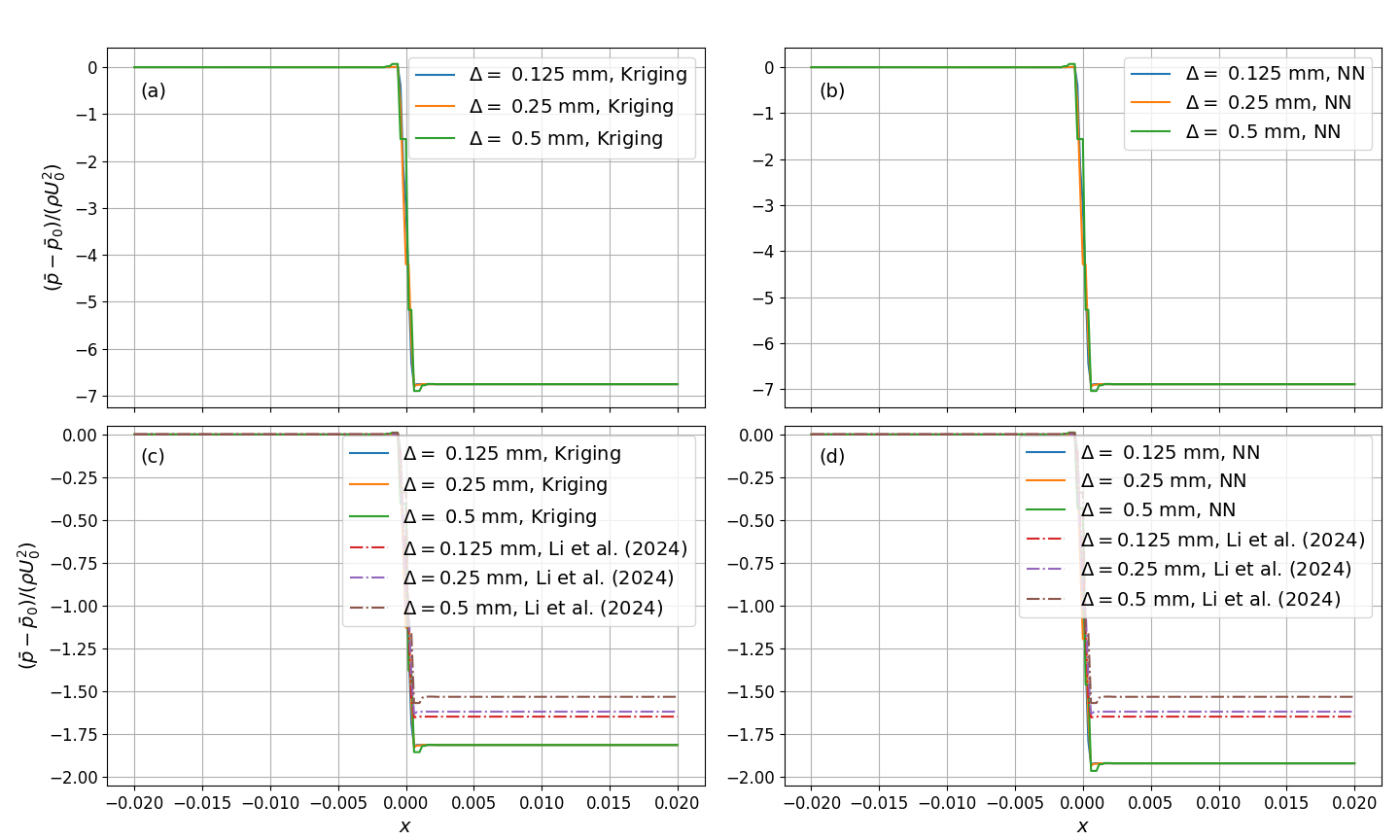}}
\caption{RANS-predicted pressure drops obtained using coarse, medium, and fine meshes. Simulation results are shown for (\textit{a}) Plate 1 with the present Kriging model, (\textit{b}) Plate 1 with the present NN model, (\textit{c}) Plate 9 with the present Kriging model and the model of Li et al.~\cite{li2024pressure}, and (\textit{d}) Plate 9 predicted by RANS with the present NN model and the model of Li et al.~\cite{li2024pressure}. The source term in the momentum equations is applied at $x=0$.} 
\label{fig:dp_RANS}
\end{figure}

\begin{figure}
\centerline{\includegraphics[angle=0,width=15.0cm]
  {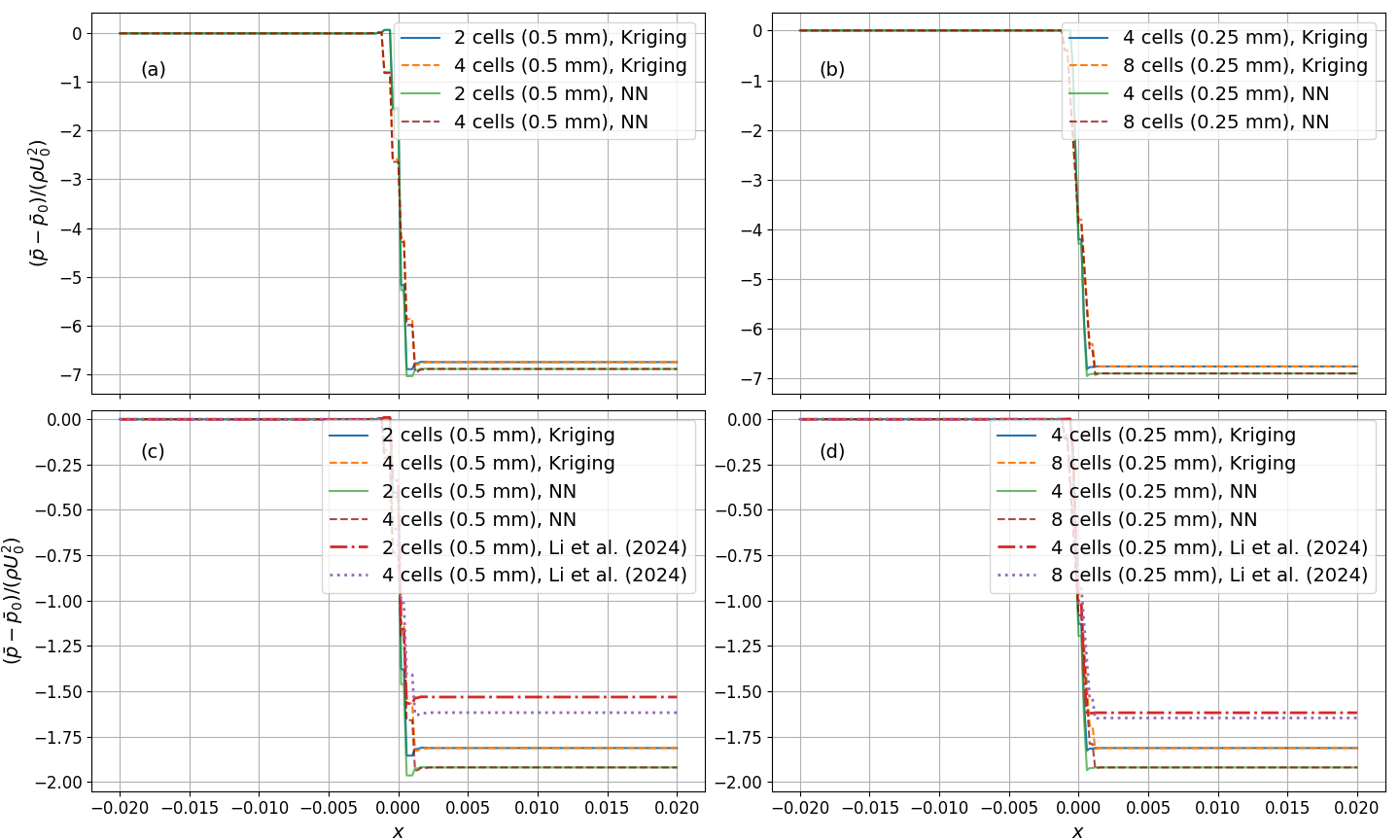}}
\caption{RANS-predicted pressure drops illustrating the effect of the number of streamwise cells used to represent the perforated plate in the simulations.  For the coarse mesh ($\Delta=0.5$ mm), simulations are performed using 2 and 4 streamwise cells to model the perforated plate, while for the medium mesh ($\Delta=0.25$ mm), 4 and 8 streamwise cells are used. Simulation results are shown for (\textit{a}) Plate 1 with the coarse mesh, (\textit{b}) Plate 1 with the medium mesh, (\textit{c}) Plate 9 with the coarse mesh, and (\textit{d}) Plate 9 with the medium mesh. Predictions using the model of Li et al.~\cite{li2024pressure} are also shown for Plate 9. The source term in the momentum equations is applied at $x=0$.} 
\label{fig:dp_RANS_oneWith4cells}
\end{figure}

Figure \ref{fig:dp_RANS} presents the pressure drops of Plates 1 and 9 predicted by the RANS simulations, in which the perforated plates are modelled using the present Kriging and NN models. Good convergence of the solutions is observed for the coarse, medium, and fine meshes. Only minor differences appear at the location of the perforated plate, i.e., at $x=0$. For comparison, RANS predictions obtained using the empirical model of Li et al.~\cite{li2024pressure} are also presented for Plate 9. As shown, when the empirical source term, $-\mu \bar{u}/K - \rho \alpha |\bar{u}| \bar{u}$, where $K$ is the permeability and $\alpha$ is the Forchheimer coefficient [48], is implemented, the RANS predictions exhibit sensitivity to mesh resolution. In contrast, the predictions obtained using the Kriging and NN models are independent of mesh resolution because the flow velocity $\bar{u}$ is not included in the source term. Figure \ref{fig:dp_RANS_oneWith4cells} further illustrates the influence of the number of streamwise cells used to represent the perforated plate in the simulations. The results clearly indicate that the solutions are insensitive to the number of streamwise cells employed for modeling the perforated plate using the Kriging and NN models. Nevertheless, a slight dependence on the number of streamwise cells is observed when the empirical correlation proposed by Li et al.~\cite{li2024pressure} is employed. This sensitivity arises because the empirical source term explicitly depends on the flow velocity, causing the predicted solution to vary with the number of streamwise cells used to represent the perforated plate. The pressure drops predicted by the RANS simulations are also summarized in Table \ref{tab:comparison} and compared with both the experimentally measured pressure drops and those predicted directly by the Kriging, NN, and Li et al.~\cite{li2024pressure} models. Good agreement is observed between the RANS predictions and the model predictions.

\renewcommand{\figurename}{Fig.}

\renewcommand{\figurename}{Fig.}

\begin{table}[width=.84\linewidth,cols=4,pos=h]
\begin{threeparttable}
\caption{Comparison of experimental pressure drops for Plates 1 and 9 with predictions from the present Kriging, NN, and Li et al.~\cite{li2024pressure} models and RANS simulations using the medium mesh.}\label{tab:comparison}
\begin{tabular*}{\tblwidth}{@{}cccccccc@{} }
\toprule
      Plate  & Exp. & Kriging & RANS-Kriging & NN & RANS-NN & Li et al.~\cite{li2024pressure} & RANS-Li et al.~\cite{li2024pressure}\\[3pt]
\midrule
       1  & 6.755 & 6.755 & 6.755 & 6.789  & 6.789 & / & / \\
       9   & 1.640 & 1.813 & 1.813 & 1.485 & 1.485 & 1.658 & 1.618\\       
\bottomrule
\end{tabular*}
\end{threeparttable}
\end{table}

\section{Concluding remarks}\label{conclusion}

A turbulent flow through a perforated plate is a common problem in a wide variety of practical applications in thermal, mechanical, chemical, civil, nuclear, ocean, and aerospace engineering \cite{li2023fluid}. Accurate modelling of the pressure loss across a perforated plate is essential for the optimal design and reliable performance assessment of thermal and fluid systems in engineering applications. In this study, two data-driven models based on Kriging and NN are proposed to predict pressure losses across perforated plates with circular perforations in turbulent flows. The models are developed using two sets of experimental data available in the literature \cite{yavuzkurt2003dependence,mery2023aerodynamic}. Here, 70\% of the available experimental data are used to train the models, while the remaining 30\% are reserved as a test dataset. The predictive performance of the proposed models is assessed and compared against widely used empirical formulae. It is found that the proposed models consistently outperform existing empirical models for most perforated plate configurations contained in the experimental datasets. Besides, the predicted pressure losses generally show good agreement with experimental measurements, demonstrating that data-driven approaches based on Kriging and NN provide a feasible framework for modelling pressure losses across perforated plates. Overall, although the Kriging model shows slightly better agreement with the experimental data than the NN model, both approaches demonstrate strong predictive capability, despite being trained on a relatively limited amount of experimental data, owing to the scarcity of measurements reported in the literature. The response surfaces of the Kriging and NN models show that the predicted pressure losses are highly sensitive to porosity at low porosity values, where the response surfaces are markedly steeper. As porosity increases, this sensitivity diminishes, resulting in progressively flatter response surfaces. In contrast, the pressure drop exhibits relatively weak sensitivity to the thickness ratio. Finally, to demonstrate the applicability of the proposed models in numerical simulations, two-dimensional channel flows are simulated using the RANS equations, in which the new pressure-loss models are implemented as source terms in the momentum equations. The RANS predictions are insensitive to mesh resolution and the number of cells used to represent the perforated plate, and they agree well with the model predictions, confirming the applicability of the proposed approaches to practical CFD simulations.

It should be noted that the proposed Kriging and NN models were developed using a limited dataset comprising 11 data points, of which 8 were used for training. Although measures such as LOOCV, L2 regularization, careful hyperparameter tuning, and a simplified NN architecture were employed to mitigate overfitting, the limited amount of training data remains a limitation of the present study. The proposed models may have limited predictive capability for cases that differ substantially from the patterns represented in the training data. Consequently, the proposed data-driven models should be regarded as interpolation tools and are intended for use only when the porosity, thickness ratio, and Reynolds number fall within the range of available experimental data. Predictions outside this range involve extrapolation and should therefore be interpreted with caution. Nevertheless, the considered ranges of porosity and thickness encompass many practical configurations, enabling the application of perforated plates in a wide variety of engineering applications, such as ventilation and HVAC systems, flow distributors, heat exchangers, silencers, filtration devices, and other industrial flow systems. For example, the proposed models are well suited for large-scale simulations involving perforated plates, such as those used for aerodynamic noise reduction in aircraft landing gears~\cite{zamponi2025noise}, where direct pore-resolved simulations are computationally prohibitive. As additional experimental data become available in the future, the models can be retrained to improve their robustness, generalization capability, and predictive accuracy.

\appendix
\section*{Appendix: Empirical formulae for pressure losses across perforated plates}
Pressure losses resulting from flow through perforated plates are often characterized by the Darcy-Forchheimer equation, expressed as:
\begin {equation}\label{pressure_gradient}
-\nabla P = \frac{\mu}{K}U_0 + \rho\alpha\lvert U_0 \rvert U_0,
\end {equation}
where $K$ is the permeability of the porous medium, $\alpha$ is the Forchheimer coefficient (also called the non-Darcy coefficient), $\rho$ is the fluid density, $U_0$ is the velocity of the incoming flow at the wind-tunnel inlet, and $\mu$ is the dynamic viscosity of the fluid. The first term on the RHS of Eq. (\ref{pressure_gradient}), known as the Darcy term, accounts for viscous pressure losses, while the second term, the Forchheimer term, represents inertial pressure losses. When the pore-level Reynolds number is sufficiently low, the Darcy term dominates over the Forchheimer term \cite{bae2016numerical, tanner2019flow, shahzad2022permeability, li2023numerical}. The pressure gradient ($-\nabla P$) on the left-hand side (LHS) of Eq. (\ref{pressure_gradient}) may alternatively be expressed using the pressure drop ($\Delta p$) across a perforated plate of thickness $\delta$, as follows
\begin {equation}\label{pressure_gradient_to_drop}
-\nabla P = \frac{\Delta p}{\delta}.
\end {equation}
The Darcy-Forchheimer drag from Eq. (\ref{pressure_gradient}) can then be expressed as \cite{lee1997modeling}
\begin {equation}\label{pressure_drop}
\frac{\Delta p D^2}{\mu \delta U_0} = \frac{D^2}{K} + \varepsilon \alpha D Re_p,
\end {equation}
where $Re_p=\rho U_0D/(\mu \varepsilon)$ is the pore-level Reynolds number. From Eq. (\ref{pressure_drop}), the normalized pressure drop across perforated plates can be expressed as
\begin {equation}\label{pressure_coef}
\frac{\Delta p}{\rho U_0^2} = \frac{\delta D}{K \varepsilon Re_p} + \alpha \delta.
\end {equation}
The first term on the RHS of Eq. (\ref{pressure_coef}) represents the contribution of the Darcy term to the normalized pressure drop. This term depends on both the plate characteristics, such as permeability, porosity, and thickness, and the flow properties, including fluid viscosity and flow velocity (or Reynolds number). The second term on the RHS corresponds to the Forchheimer contribution, which depends solely on the plate characteristics, including porosity and thickness.

Although Eq. (\ref{pressure_coef}) provides a mathematical framework for describing pressure loss across perforated plates, the permeability $K$ and Forchheimer coefficient $\alpha$ are yet to be determined. Based on numerical simulations of laminar flows ($Re_p<25$), Bae \& Kim \cite{bae2016numerical} developed the following model to estimate $K$ and $\alpha$ for perforated plates:
\begin {equation}\label{parameters_BaeKim2016}
K = \frac{\varepsilon D^2 \delta}{32\delta+15D}, \:\:\: \alpha = \frac{3(1-\varepsilon)}{4\varepsilon^2\delta}.
\end {equation}
This model has been validated using the experimental data of Doerffer \& Bohning \cite{doerffer2000modelling}. For a detailed comparison between the model predictions and the measurements, the reader is referred to Ref. \cite{bae2016numerical}.

Besides, several models for pressure losses across perforated plates in turbulent flow regimes have been proposed by researchers such as Idelchik \cite{idelchik1986handbook, idelchik1994handbook}, Miller \cite{miller1990internal}, Kast \cite{kast2010pressure}, Holt et al. \cite{holt2011cavitation}, and Li et al. \cite{li2024pressure}. In these regimes, the Forchheimer term strongly dominates over the Darcy term, allowing the Forchheimer coefficient $\alpha$ to be used to calculate the Forchheimer contribution, which is approximately equal to the total pressure loss. For perforated plates of finite thickness at high Reynolds numbers, Idelchik \cite{idelchik1994handbook} proposed a formula of the following form:
\begin {equation}\label{Idelchik1994}
\alpha = \frac{1}{2\varepsilon^2\delta}\left(0.5+0.24\sqrt{1-\varepsilon}(1-\varepsilon)+(1-\varepsilon)^2\right).
\end {equation}
Similarly, Kast \cite{kast2010pressure} suggested the following formula:
\begin {equation}\label{Kast2010}
\alpha = \frac{1}{2\varepsilon^2\delta}\left((\frac{1}{C}-1)^2+(1-\varepsilon)^2\right),
\end {equation}
where $C$ is a coefficient that depends on the plate porosity and can be expressed as
\begin {equation}
C = 0.6 + 0.4\varepsilon^2.
\end {equation}
Miller \cite{miller1990internal} proposed the following formula:
\begin {equation}
 \alpha = C_0\frac{(1-C_c\varepsilon)^2}{2C_c^2\varepsilon^2\delta},
\end {equation}
where $C_0$ is a coefficient that depends on the thickness ratio $\delta/D$, and $C_c$ is the jet contraction coefficient, which depends on the porosity $\varepsilon$. According to Fratino \cite{fratino2000hydraulic}, $C_0$ can be calculated using the following empirical expression, valid for $0.1<\delta/D<3$:
\begin {equation}
C_0 = 0.5 + \frac{0.178}{4(\frac{\delta}{D})^2+0.355},
\end {equation}
while $C_c$ is given by
\begin {equation}
C_c = 0.596 + 0.0031\exp(\frac{\sqrt{\varepsilon}}{0.206}).
\end {equation}
Holt et al. \cite{holt2011cavitation} proposed the following piecewise function for the Forchheimer coefficient:
\begin {equation} 
\label{convective_term_BCD}
\alpha = 
    \begin{cases}
        \frac{1}{2\delta}\left(2.9-3.79\frac{\delta}{D}\varepsilon^{0.2}+1.79\left(\frac{\delta}{D}\right)^2\varepsilon^{0.4}\right)K_{LA}  & \frac{\delta}{D}\varepsilon^{0.2}<0.9 \\
        \frac{1}{2\delta}\left(0.876+0.069\frac{\delta}{D}\varepsilon^{0.2}\right)K_{LA}  & \hfill \break \frac{\delta}{D}\varepsilon^{0.2}>0.9,
    \end{cases}
\end {equation}
where $K_{LA}$ is the pressure-loss coefficient of a single-hole orifice, estimated using a theoretical model for reattached flows, as follows:
\begin {equation}
K_{LA} = 1 - \frac{2}{\varepsilon} + \frac{2}{\varepsilon^2}\left(1-\frac{1}{C_c}+\frac{1}{2C_c^2}\right),
\end {equation}
where the jet contraction coefficient $C_c$ is set to 0.72 \cite{malavasi2012pressure}.
Recently, Li et al. \cite{li2024pressure} proposed a pressure-drop model for perforated plates applicable to both laminar and turbulent flow regimes, as follows:
\begin {equation}\label{parameters_improved}
K = \frac{\varepsilon D^2 \delta}{32\delta+15D}, \:\:\: \alpha = \frac{9}{40\varepsilon^2\delta}\left[6\left(\frac{\delta}{D}\right)-5\left(\frac{\delta}{D}\right)^2\right].
\end {equation}
The model proposed by Li et al. \cite{li2024pressure} has been validated against existing numerical simulations~\cite{shahzad2022permeability} in the laminar flow regime and experimental measurements~\cite{yavuzkurt2003dependence} in the turbulent regime, demonstrating good agreement.

\printcredits

\section*{Declaration of Competing Interest}
The author declares that he has no known competing financial interests or personal relationships that could have appeared to influence the work reported in this paper.

\section*{Funding Source}
This work is supported by the project Innovative Design of Installed Airframe Components for Aircraft Noise Reduction ("INVENTOR", European Union’s Horizon 2020 Research and Innovation Programme, under Grant Agreement N$^\circ$ 8605383), and by the Horizon Europe project Scale-resolving Simulations for Innovations in Turbomachinery Design ("Sci-Fi-Turbo") under Grant Agreement No. 101138080.

\section*{Acknowledgments}
The author would like to thank the partners in the European projects for their valuable and helpful discussions. The computations were enabled by the computer resources at the Chalmers Centre for Computational Science and Engineering (C3SE) provided by the the National
Academic Infrastructure for Supercomputing in Sweden (NAISS). 

\section*{Data Availability}
Data will be made available on request.

\bibliographystyle{elsarticle-num}

\bibliography{cas-refs}


\end{document}